\documentclass[conference]{IEEEtran}
\IEEEoverridecommandlockouts
\usepackage{url}
\usepackage{cite}
\usepackage{amsmath,amssymb,amsfonts}
\usepackage{algorithmic}
\usepackage{graphicx}
\usepackage{textcomp}
\usepackage{xcolor}
\usepackage{csquotes}
\usepackage{epstopdf}
\usepackage{tikz}
\usepackage{circuitikz}
\usepackage{subfigure}
\usepackage{multirow}
\usepackage{url}
\usepackage{siunitx}
\usepackage{amsthm}

\usepackage{booktabs}
\usepackage{tabularx}

\def\BibTeX{{\rm B\kern-.05em{\sc i\kern-.025em b}\kern-.08em
    T\kern-.1667em\lower.7ex\hbox{E}\kern-.125emX}}

\def\BibTeX{{\rm B\kern-.05em{\sc i\kern-.025em b}\kern-.08em
		T\kern-.1667em\lower.7ex\hbox{E}\kern-.125emX}}
\newcommand{\R}{\mathbb{R}}
\newcommand{\Z}{\mathbb{Z}} 
\newcommand{\tran}{^{\mbox{\scriptsize T}}} 
\newcommand{\C}{\mathbb{C}}

\newcommand{\com}[1]{\textcolor{black}{#1}}
\newcommand{\cf}{\textit{cf.~}}

\begin{document}

\title{On Musical Onset Detection via the S-Transform\\
}

\author{\IEEEauthorblockN{Nishal Silva}
	\IEEEauthorblockA{\textit{Dept. of Eng. and Mathematics} \\
		\textit{Sheffield Hallam University}\\
		Sheffield, UK \\
		b3047941@my.shu.ac.uk}
	\and
		\IEEEauthorblockN{Chathuranga Weeraddana}
	\IEEEauthorblockA{\textit{Dept. of Electronic and Telecomm. Eng. } \\
		\textit{University of Moratuwa }\\
		Moratuwa, Sri Lanka \\
		chathurangaw@uom.lk}
	\and
	\IEEEauthorblockN{Carlo Fiscione}
	\IEEEauthorblockA{\textit{Dept. of Networks and Systems Eng.} \\
		\textit{KTH Royal Institute of Technology}\\
		Stockholm, Sweden \\
		carlofi@kth.se}
}

\maketitle

\begin{abstract}
Musical onset detection is a key component in any beat tracking system. Existing onset detection methods are based on temporal/spectral analysis, or methods that integrate temporal and spectral information together with statistical estimation and machine learning models. In this paper, we propose a method to localize onset components in music by using the S-transform, and thus, the method is purely based on temporal/spectral data. Unlike the other methods based on temporal/spectral data,  which usually rely short time Fourier transform (STFT), our method enables effective isolation of crucial frequency subbands due to the frequency dependent resolution of S-transform. Moreover, numerical results show, even with less computationally intensive steps, the proposed method can closely resemble the performance of more resource intensive statistical estimation based approaches.


\end{abstract}


\section{Introduction}\label{sec:Introduction}
When a human hears music, an action which is almost subconscious is the rhythmic tapping of the foot. These taps are consistent with the \textit{beat} of the music and is measured in beats-per-minute (\texttt{}{bpm}). The process of detecting beat locations in a music is called \emph{beat tracking}. Beat tracking is a vital step in many studio and live music applications: for example, when a DJ should perform beat matching to play two songs successively. Beat matching is the adjustment of the tempo of one or multiple songs so that their beat locations overlap each other when played simultaneously. The same applies for an audio engineer whenever two instrument tracks are to be played in unison. In this case the audio engineer needs to know the beat locations in both tracks to create a smooth playthrough. 

Beat of a music is maintained by a rhythm instrument. A beat usually corresponds to a rapid and unpredictable change in the underlying music signal. Therefore, a primary step in any beat tracking algorithm is to represent such changes, which is referred to as the \emph{beat causing onsets} (BCO). However, isolating BCOs among others can be challenging. 

Existing onset isolation (detection) algorithms, based on temporal and spectral analysis, do not usually yield good results when the beat of a music is not prominent. A primary cause of this is the masking off of important BCO components. Therefore, exploring generalized mechanisms for BCO detection in music, is important in theory, as well as in practice, and therefore deserve investigation. Blending the existing temporal and spectral analysis methods with statistical estimation techniques yields more promising results, however, at the expense of significant computational complexity. Integrating temporal and spectral data of music with machine learning techniques (e.g., neural networks) is apparently the best among others. Such algorithms always rely on a substantial training phase in advance, in order to yield promising results. 

In this paper, we propose a method which relies on the S-transform~\cite{stockwell1} for BCO detection. Unlike the existing methods based on statistical estimation techniques, our method does not rely on any \emph{a priori} information of the underlying music. Moreover, unlike the state-of-the art machine learning algorithms, the proposed method does not require a training dataset. The proposed algorithm can be considered as a graceful trade-off between the performance and the computational complexity and resources required. 

The choice of the S-transform, among other time-frequency representations (TFR), is motivated by the following:

\begin{enumerate}
\item The beat causing onsets are usually created by instruments with relatively lower frequencies~\cite{marxer1},
\item S-transform provides a good concentration at lower frequencies~\cite{stockwell1}.
\item S-Transform uses a frequency-dependent window dilation, which results a frequency-dependent resolution~\cite{stockwell1}.
\end{enumerate}

The first two points enable one to extract the power of rhythm instruments effectively. The last point plays a key role in the sense that, unlike the STFT, S-transform is not required to know the window size a priori. This facilitates, irrespective of the underlying frequencies of the rhythm instruments, a general implementation of proposed algorithms.

The rest of the paper is organized as follows. In Section~\ref{sec:Literature}, we give a literature overview. Section~\ref{sec:ProposedMethod} discusses our proposed algorithms for BCO detection. In
Section~\ref{sec:results}, numerical results are presented. Section~\ref{sec:Conclusion} concludes the paper.


\section{Literature}\label{sec:Literature}

Several works have been investigated on BCO detection in music~\cite{klapuri1,scheirer1,ellis1,mcfee1,laroche1,alonso1,stark1,wu1,goto1,shiu1,cemgil1,zapata1,degara1,fiocchi1,bock1,bock2,bock3,klapuri4,dixon1,davies1,duxbury1}. These can be split into methods based on temporal/spectral analysis, and more sophisticated methods which blend temporal/spectral data together with statistical estimation techniques and machine learning techniques. Temporal and spectral analysis methods generate a \emph{time series}, usually called the \emph{onset envelope function} (OEF), which contains information of the locations of BCOs. The OEF is then used to compute the underlying~bpm~\cite{ellis1}. Methods based on statistical estimation and machine learning techniques rely on more resources, in addition to the pure temporal and spectral data, for locating BCOs, e.g., \emph{a priori} information of the underlying music, training data sets~\cite[\S~4]{fiocchi1}.

Temporal analysis methods split the signal into frequency bands, for which amplitude envelopes are calculated and summed to obtain an OEF \cite{klapuri1,scheirer1}. Spectral analysis methods take into account, the change in spectral energy. These methods usually compute some form of a time-frequency representation (TFR), where the STFT is most common~\cite{dixon1,davies1,klapuri1,ellis1,wu1}. Different scaling methods such as \emph{the Mel} \cite{ellis1,mcfee1} and \emph{the square~root}~\cite{laroche1} are used to avoid low amplitude components from being masked off. Either the summation~\cite{ellis1,laroche1}, the median~\cite{mcfee1}, or the mean~\cite{alonso1} is computed of the first order difference for each time bin to obtain an \emph{OEF}.

A common limitation of the spectral analysis methods mentioned above is the poor detection of BCOs if the rhythm is less pronounced. This is due to masking off of BCO components of interest, or because spectral changes constituting to BCOs have not been identified accurately. This is the case in most classical, opera, soft pop and instrumental music~\cite{duxbury1}. In addition, the designs can be very sensitive to the algorithm parameters, e.g., window length~\cite{scargle1}.

The authors of \cite{gouyon1} presents a comparison between several onset detection methods which were submitted to the ISMIR 2006 competition~\cite{ismir}. The methods discussed include the works presented by \cite{alonso1,dixon1,klapuri4,scheirer1} and several others. The authors show that the method proposed by~\cite{klapuri4}, which maneuvers temporal/spectral data, together with statistical estimation techniques, outperforms other methods by a considerable margin.

Methods such as~\cite{fiocchi1,bock1,bock2,bock3} uses a machine learning based approach where there is no computation of an OEF. Based on recent results of ISMIR~\cite{ismir}, the research conducted in~\cite{bock1,bock2,bock3} appears to be the best among others. However, for machine learning algorithms, usually the existence of a reasonable training data set is necessary to achieve better accuracies.

\begin{figure}[t]
\centering
{\includegraphics[width=0.47\textwidth]{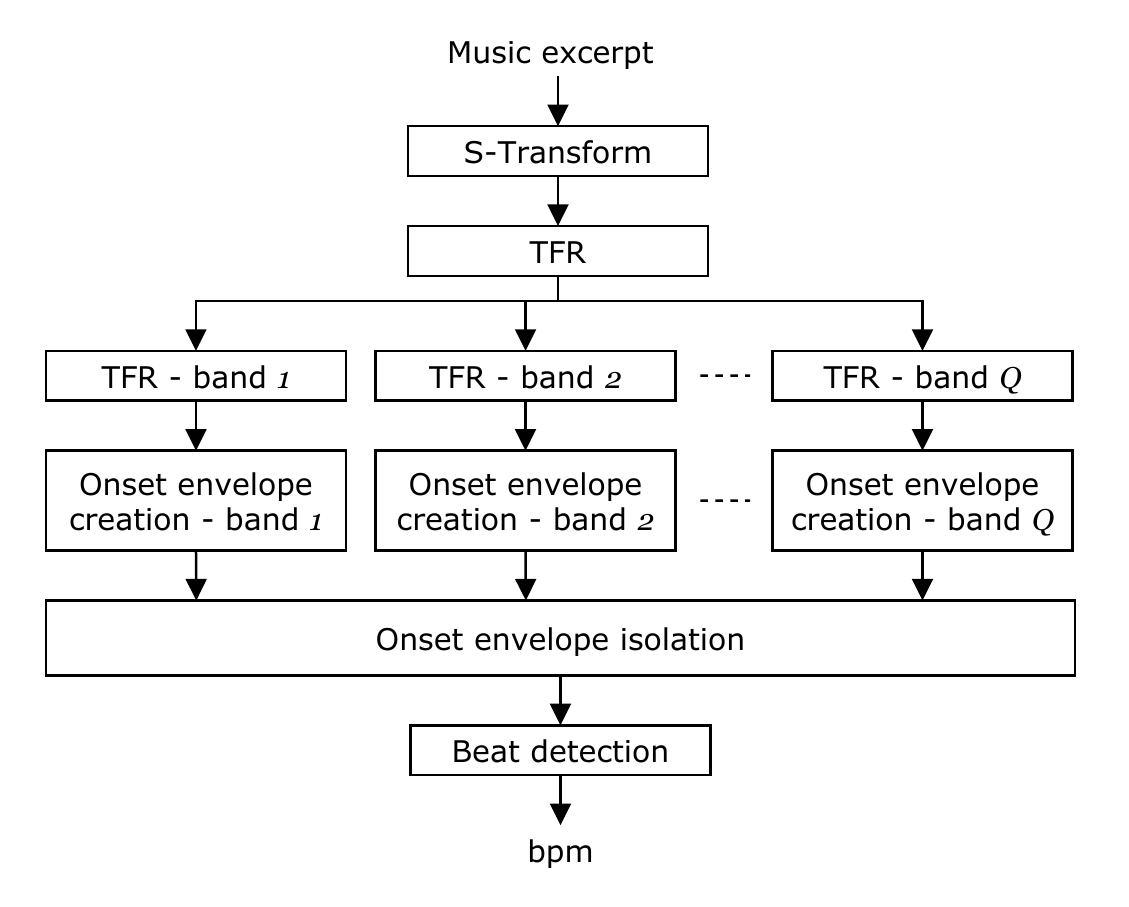}}
\caption{Black diagrams of Proposed Method.}
\label{fig:Existing-Vs-Proposed} 
\vspace{-0mm}
\end{figure}

\section{Proposed Method, An Overview}\label{sec:ProposedMethod}

The proposed method is based on the discrete S-transform~\cite[\S~III]{stockwell1}. Moreover, the overall method is divided into two sections;
\begin{enumerate}
	\item Onset envelopes by band spitting.
	\item Onset envelope isolation.
\end{enumerate}

Recall that the existing methods rely on a single STFT-TFR followed by an associated onset envelope for beat detection. Intuitively, to get the benefits of frequency dependent resolution of S-transform, it is suggestive to split the TFR into several bands and to process different subbands separately~\cite{klapuri1,scheirer1,klapuri4}. Such a splitting and a processing can avoid or at least minimize the masking off and suppression of desired BCO information from undesired spectral information. Thus, we first consider a band splitting followed by an onset envelope computation for each band (Figure~\ref{fig:Existing-Vs-Proposed}). Note that, of the several onset envelopes present, the beat information may be encoded in some, depending on the rhythm instrument used. The challenge is then to pick the `best' one that encodes the BCOs of the underlying music. This is the second stage of our proposed method, in particular the onset envelope isolation, see Figure~\ref{fig:Existing-Vs-Proposed}.  

In the sequel, we discuss in more detail, the computation of onset envelopes by band splitting [\textit{cf.}~\S~\ref{sec:Band Partitioning}] and onset envelope isolation [\textit{cf.}~\S~\ref{sec:Band Isolation}].

\section{Onset Envelopes by Band Spitting}\label{sec:Band Partitioning}

Let us first outline the proposed algorithm for onset envelope computation. We assume that the musical excerpt is provided in mono format.

\noindent\rule{\columnwidth}{0.3mm}
\\
\textbf{\emph{Algorithm~1}}\\ 
\noindent\rule{1\columnwidth}{0.3mm}

\noindent\textbf{Input:} 
\begin{itemize}
	\item Mono audio file, $\{x[n] \}_{n=0}^{N-1}$.
	\item Downsamling factor $D$, a positive even integer.
	\item Subband size, $K$ such that $\lfloor{(N-1)/D}\rfloor=2QK-1$ for some positive integer $Q$. 
\end{itemize}

\noindent\textbf{Steps:}
\begin{enumerate}
	
	\item Downsamling:
	\begin{equation}\nonumber
	y[n] = x[nD], \quad n=0,\ldots,M-1,
	\end{equation}
	where $M = 1+\lfloor{(N-1)/D}\rfloor$.
	\item Compute $M$-Discrete Fourier Transform $\{Y[k] \}_{k=0}^{M-1}$ of $\{y[n] \}_{n=0}^{M-1}$, where
	\begin{equation}\nonumber
	\label{eq:DFT}
	Y[k] = \frac{1}{M} \sum\limits_{n=0}^{M-1} y[n] \exp{\left(\displaystyle-\frac{j2\pi nk}{M}\right)}.
	\end{equation}
	
	\item Compute Discrete S-Transform matrix $F\in\C^{(M/2)\times M}$, whose $(p,n)$-th element is given by:

	\begin{equation}\nonumber
	F[p, n]=\begin{cases}
	\displaystyle \sum\limits_{m=0}^{M-1} Y[m+n] \exp{\left(\frac{j2\pi mp}{N}-\frac{2\pi^2 m^2}{n^2}\right)}, & \\
	\qquad \qquad \qquad \quad  \quad \ \ \qquad \qquad \quad \text{if $n\neq 0$} & \\
	\displaystyle\frac{1}{M}\sum\limits_{m=0}^{M-1}y[m], \qquad \qquad \quad \quad  \ \ \ \text{otherwise}, & 
	\end{cases}
	\end{equation}
	where $n=0,\ldots,M-1$ and $p=0,\ldots,M/2-1$. Define $S\in\R^{(M/2)\times M}_+$ as follows:
    \begin{equation}\nonumber
    S(p,n) = |F(p,n)|, \quad \forall p,n.
    \end{equation}

	\item Split $S$ by rows,
	\begin{equation}\nonumber
	S = [S_1\tran \ S_2\tran \ \cdots \ S_Q\tran]\tran,
	\end{equation}
	with $S_i\in \R^{K\times M}$ representing $i$-th block of $S$.

	\item For each block $S_i$, compute the mean (over rows) $r_i\in\R^{M}$, i.e., 
	\begin{equation}\nonumber
	r_i = K^{-1}S_i\tran 1, 
	\end{equation}
	where $1\in\R^{K}$ is a $K$-vecor with all ones.
\end{enumerate}

\noindent\textbf{Output:} 
\begin{itemize}
	\item Onset envelopes: return $r_i\in\R^M$ associated with subband~$i$, $i=1,\ldots,Q$.
\end{itemize}	

\vspace{-2mm}
\noindent\rule{1\columnwidth}{0.3mm}\vspace{-0mm}

Algorithm starts with a sampled musical excerpt denoted by the sequence $\{x[n]\}_{n=0}^{N-1}$. Note that, the smaller $N$ or the duration $T$ of the musical excerpt is, the lesser the computational burden of the algorithm. Therefore, the duration $T$ can be chosen intelligently for efficient implementation of the algorithm. Note that, the tempo of a music can usually range from $60$~bpm to $240$~bpm~\cite{apel1}. Therefore, $T$ can be on the order of few seconds to extract useful beat information. For example, even in the worst-case, i.e., when the musical excerpt is of $60$~bpm, a $T=4$ second musical excerpt can be used to capture $4$ beats for further processing. 

The downsampling factor $D$ also plays a key role for efficient implementation of the algorithm [\textit{cf.} step~(1)]. In other words, the larger $D$ is, the smaller $M$, and therefore, the lesser computational burden of the algorithm [\textit{cf.} step~(2), (3)]. A better choice for~$D$ can be argued by considering the frequencies of rhythm instruments. Note that the frequencies of rhythm instruments' typically range from $32$Hz to $512$Hz~\cite{watkinson1}. Thus, sampling is to be done at a rate no smaller than $1024$Hz to avoid aliasing. Therefore, for a musical excerpt sampled at a rate $f_s=44100$Hz~\cite{watkinson1}, $D=40$ corresponds to a sampling frequency $1102.5$Hz ($\geq 1024$Hz) and $M=4410$ samples in a $T=4$~s period.

The idea of band splitting is essentially to extract the potentials of S-transform in a frequency dependent resolution. Thus, the choice of $K$ is to be such that it is large enough to hold a sufficient spectral energy concentration to emphasize BCO~s~(if any). On the other hand, $K$ should be small enough to minimize the masking off of important BCO information~(if any) from spectral contents within the subband itself. Numerical experiences suggest that a $K$ on the order of $200$ for a $T=4$~s period, or in other words, a subband width on the order of $50$Hz is a good~choice. Note that the subbands are indexed by $1,\ldots,Q$ for simplicity.

\begin{figure}[t]
\begin{center}
	\begin{tikzpicture}	
	
	\draw[gray, thick] (-2,-2	) -- (2,-2	);
	\draw[gray, thick] (-2,-1.5	) -- (2,-1.5	);
	\draw[gray, thick] (-2,-1	) -- (2,-1);
	\draw[gray, thick] (-2,-0.5	) -- (2,-0.5	);
	\draw[gray, thick] (-2,0	) -- (2,0	 );
		
	\draw[gray, thick] (-2,-2)  -- (-2,0 );
	\draw[gray, thick] (2,-2)  -- (2,0 );
	
	\draw[gray, thick, ->] (-2,-2	) -- (3,-2	 );
	\draw[gray, thick, ->] (-2,-2	) -- (-2,0.4	 );
	
	\node (sQ)  at (0,-0.25)[circle] {\small $S_Q^T$};
	\node (sd)  at (0,-0.65)[circle] {\small $\vdots$};
	\node (s2)  at (0,-1.25)[circle] {\small $S_2^T$};
	\node (s1)  at (0,-1.75)[circle] {\small $S_1^T$};

	\node (m0) at (-2, -1.9) [anchor=east] {\small $f = 0$ Hz};
	\node (m0) at (-2, 0) [anchor=east] {\small $551.25$ Hz};
	
	\node (m0) at (2, -1.8) [anchor=west] {\small $p = 0$};
	\node (m0) at (2, 0) [anchor=west] {\small $M/2$};
	
	\node (m0) at (-1.8, -2.3) [anchor=east] {\small $t=0$};
	\node (m0) at (1.8, -2.3) [anchor=west] {\small $T$};
	
	\node (m0) at (-1.8, -2.6) [anchor=east] {\small $n=0$};
	\node (m0) at (1.75, -2.6) [anchor=west] {\small $M$};

	\end{tikzpicture}
\end{center}
\caption{Splitting of discrete S-transform matrix $S\in\R^{(M/2)\times M}$}
\label{fig:s-transform-split}
\end{figure}
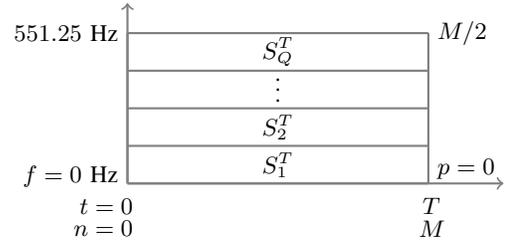

A concise depiction of our considered TFR, in particular, the \emph{absolute discrete S-transform matrix} $S\in\R^{(M/2)\times M}$ is shown in Figure~\ref{fig:s-transform-split}, together with the considered splitting. Note that the TFR is plotted only for the range $0\textrm{Hz} \leq f \leq 551.25\textrm{Hz}$ and $0~\textrm{s} \leq t \leq T~\textrm{s}$, because the upper frequency band $551.25\textrm{Hz} < f \leq 1102.5\textrm{Hz}$ is just a repetition of $S$. 

After having determined $S$ and its splitting [\textit{cf. step (3), (4)}], step~(5) computes the onset envelopes of each subband. The output of the algorithm is the onset envelopes for each subband, which is used by the onset envelope isolation stage.

\section{Onset Envelop Isolation}\label{sec:Band Isolation}

Given onset envelopes $r_i\in\R^M, i=1,\ldots,Q$, the task of the isolation stage is to choose \emph{one} envelope that can potentially encode the BCO information. To this end, the key idea is to associate each $r_i$, with a real number $b_i$, so that, the bigger $b_i$ is, the higher the likeliness of $r_i$ carrying BCO information. Let us first outline the algorithm.

\noindent\rule{\columnwidth}{0.3mm}
\\
\textbf{\emph{Algorithm~2}}\\ 
\noindent\rule{1\columnwidth}{0.3mm}

\noindent\textbf{Input:} 
\begin{itemize}
	\item Onset envelopes: $r_i\in\R^M, i=1,\ldots,Q$.
	\item Local maxima (peak) separation  $n_p$.
    \item Threshold steps $H$.
    \item Isolation accuracy level $\epsilon>0$.
\end{itemize}

\noindent\textbf{Steps:}

For each $i\in\{1,\ldots,Q\}$,
\begin{enumerate}
	\item Normalization: compute $\tilde{r}_i$ as $\tilde{r}_i = {r_i}/{||r_i||_\infty}$, where $||\cdot||_\infty$ is the $\ell_\infty$ norm.
    
	\item Upper envelope computation: Determine the upper envelope $u_i\in\R^M$ by using \emph{cubic} spline interpolation over local maxima of $\tilde r_i$ separated by at least $n_p$ samples~\cite[\S~IV]{boor1}.
    \item Centering: Compute $\hat{r}_i$ as
       \begin{equation}\nonumber
    	\hat{r}_i = [\tilde r_i - (1\tran u_i/M) 1]_+,
    \end{equation}
where $1\in\R^M$ is a $M$-vector with all ones and $[x]_+$, is the projection of $x$ onto $\R^M_+$~\footnote{That is the vector obtained by taking the nonnegative part of each component of $x$ and replacing each negative component with $0$.}.

	\item Thresholding and clustering: Divide equally, the range $\mathcal{H}_i=[0,\max(\hat r_i)]$ into $H$ segments indexed by~$\{1,\ldots,H\}$.    
    \vspace{1em}
    
    For each segment $j\in\{1,\ldots,H\}$ 
    \begin{enumerate}
    \item Let threshold $h=l_j$, the lower level of segment~$j$.
    \item Let $\mathcal{I}=\{k \ | \ (\hat r_i)_k\geq h\}$, the set of indexes whose associated components are larger than or equal to the threshold~$h$.  
       \item Determine the set partition $\{\mathcal{I}_m\}_{m=1}^{M_i}$ of $\mathcal{I}$ such that, $\mathcal{I}_m\cap\mathcal{I}_{\bar m}=\emptyset$ $\forall m,\bar m$ and the elements of any set are \emph{consecative}. 
    \item Let $\{\bar I_m\}_{m=1}^{M_i}$ be the ordered sequence, where $I_m$ is  of  mean of the elements of $\mathcal{I}_m$.
    \item Define $c_{ij}\in\R^{M_{i}-1}$ as follows:
    \begin{equation*}
    c_{ij} = [I_{12},I_{23},\ldots,I_{(M_i-2)(M_i-1)},I_{(M_i-1)(M_i)}]\tran,
    \end{equation*}
    where $I_{mn}=I_n-I_m$ and let $v_{ij}=(1\tran c_{ij})/||c_{ij}||_2$.

    \end{enumerate}
 
\item Define $b_i\in\R$ as follows:
\begin{equation*}
b_i = \max_{j\in\{1,\ldots,H\}}v_{ij}.
\end{equation*}
   
\end{enumerate}

\noindent\textbf{Output:}
\begin{itemize}
\item  Onset envelope isolation:
\begin{equation*}
\mathcal{I}^\star = \{i \ | \ |1-b_i|\leq \epsilon, \ i\in\{1,\ldots,Q\}\}.
\end{equation*}
\item If $\mathcal{I}^\star=\emptyset$, return an exception \texttt{Isolation Failure}, Otherwise return $r_{i^\star}$, where the partition index $i^\star\in\mathcal{I}^\star$.
\end{itemize}

\vspace{-2mm}
\noindent\rule{1\columnwidth}{0.3mm}\vspace{-0mm}

\begin{figure}[ht]
	\centering
	\includegraphics[width=0.9\columnwidth]{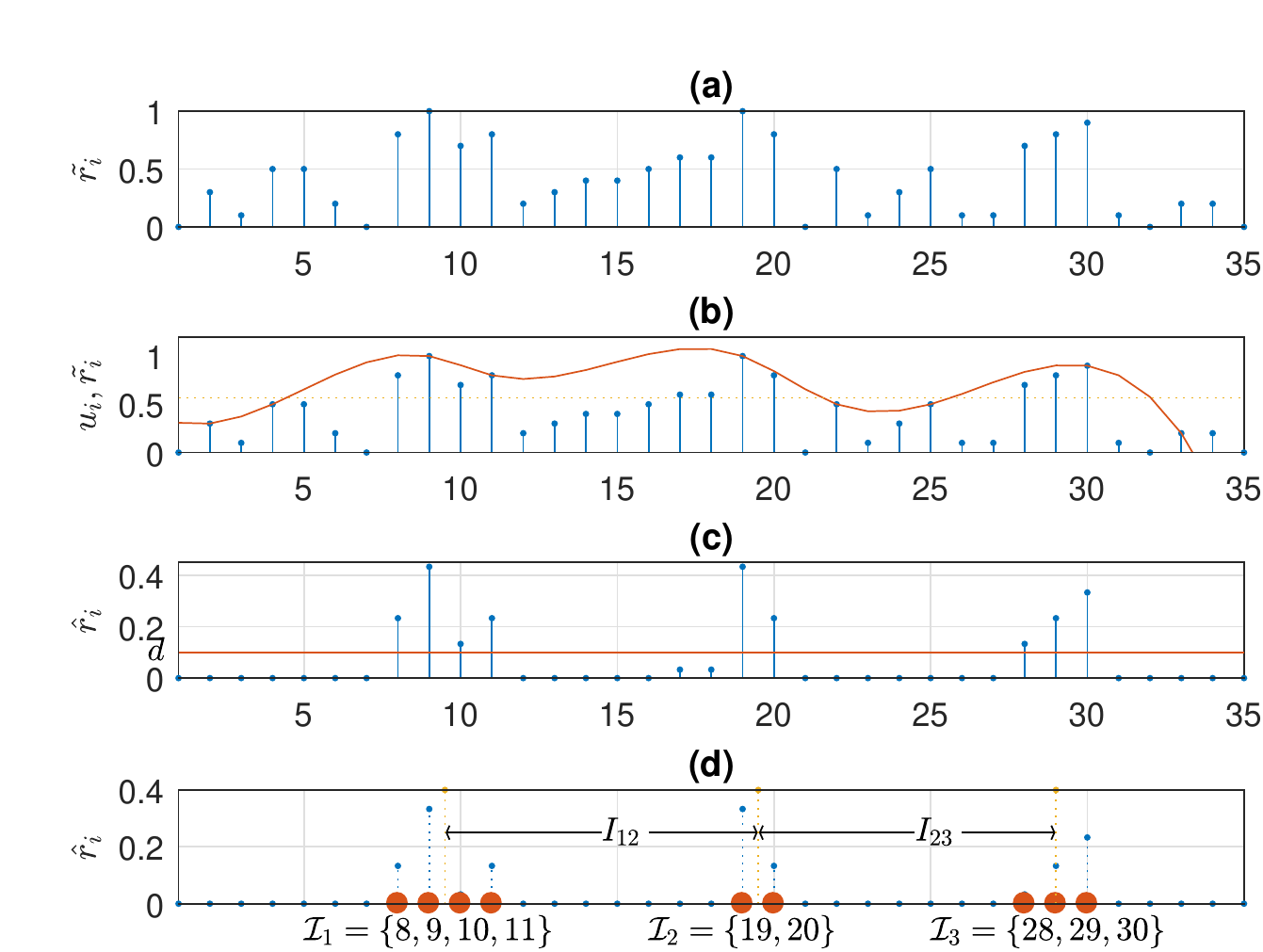}
	\caption{Signatures of split frequency bands}
	\label{Alg2-Illus}	
\end{figure}

The first step is a preconditioning step, where $r_i$ is normalized to yield $\tilde r_i$. For an illustration, see \com{Figure~\ref{Alg2-Illus}-(a)}. It is reasonable to assume that most of the relatively lower level amplitudes of $\tilde r_i$ do not carry BCO information. Therefore, we consider only the amplitudes of $\tilde r_i$ above some level. More specifically, the level is chosen to be the \emph{mean} [\com{Figure~\ref{Alg2-Illus}-(b)}, dotted curve] of the \emph{upper envelope} $u_i$ [\com{Figure~\ref{Alg2-Illus}-(b)}, solid curve] determined at step~(2). Step~(3) removes the mean aforementioned from $\tilde r_i$ to yield~$\hat r_i$,~\cf{\com{Figure~\ref{Alg2-Illus}-(c)}}. 

Note that the upper envelope $u_i$ in step~(2), computed by using cubic spline interpolation corresponds to some local maxima~\footnote{We say $k\in\Z$ is a local maximum of $x\in\R^M$ whenever $(x)_{k-1}<(x)_{k}<(x)_{k+1}$, where $(x)_k$ represents the $k$-th component of $x$.} of $\tilde{r}_i$ whose separation is at least $n_p\in\Z$ samples. For example, \com{Figure~\ref{Alg2-Illus}-(b)} shows $u_i$ of $\tilde r_i$ in \com{Figure~\ref{Alg2-Illus}-(a)} for~$n_p=1$. 


Step~(1), (2), as well as (3) of the algorithm correspond to preconditioning of the input $r_i$. In contrast, step~(4) is the key for envelope isolation, which capitalizes on a clustering of components of $\hat r_i$ by using a thresholding mechanism. To see this, first suppose the range of frequencies of the underlying rhythm instrument overlaps with subband $i\in\{1,\ldots,Q\}$. Thus, there is a high potential that $\hat r_{i}$ contains nonzero components, which correspond to the BCOs. In addition, their neighboring components can also be nonzeros due to the \emph{spectral leakage} caused by windowing. As a result, $\hat r_i$ can resemble a sequence as shown in \com{Figure~\ref{Alg2-Illus}-(c)}, where there are  clusters of nonzero components (\emph{nonzero clusters}) that are separated by clusters of zero components (zero clusters). For example, $\hat r_i$ in  \com{Figure~\ref{Alg2-Illus}-(c)} has $3$ nonzero clusters. Because of the periodicity of BCOs, the `distance' between consecutive pairs of nonzero clusters should be the same. However, for any subband $\bar i \neq i$, the characteristics of the nonzero clusters mentioned above, do \emph{not} apply. This is indeed the key to isolate subband $i$ from others. 

Steps~(4)-a to (4)-e, correspond to clustering and the distance computation between consecutive pairs of nonzero clusters of $\hat r_i$. First, a threshold $h$ is given, \cf step~(4)-a and \com{Figure~\ref{Alg2-Illus}-(c)}. Then components of $\hat r_i$ which are greater than or equal to $h$ is isolated into $\mathcal{I}$, \cf{ step~(4)-b}. For example, \com{Figure~\ref{Alg2-Illus}-(c)} shows that $\mathcal{I}=\{8,9,10,11,19,20,28,29,30\}$. Step~(4)-c partitions $\mathcal{I}$ into subsets $\{\mathcal{I}_m\}_{m=1}^{M_i}$, where each subset corresponds to a nonzero cluster. For example, from \com{Figure~\ref{Alg2-Illus}-(d)}, we have $M_i=3$ subsets (one for each nonzero cluster), denoted $\mathcal{I}_1,\mathcal{I}_2$, and $\mathcal{I}_3$, where $\mathcal{I}_1=\{8,9,10,11\}$, $\mathcal{I}_2=\{19,20\}$, and $\mathcal{I}_3=\{28,29,30\}$. Step~(4)-d computes the center of gravity of each subset, denoted $\{I_m\}_{m=1}^{M_i}$. Particularized to our example, we have $I_1=9.5$, $I_2=19.5$, and $I_3=29$, \cf{\com{Figure~\ref{Alg2-Illus}-(d)}}. The distance between consecutive pairs of nonzero clusters are simply given by the $(M_i-1)$-vector $[I_{12},I_{23},\ldots,I_{(M_i-2)(M_i-1)},I_{(M_i-1)(M_i)}]\tran$, \cf{step~(4)-e}. This is illustrated in \com{Figure~\ref{Alg2-Illus}-(d)}, where the distance between nonzero cluster~$1$ and $2$ is $I_{12}$ and that of nonzero cluster~$2$ and $3$ is $I_{23}$. Finally, recall that the `distance' between consecutive pairs of nonzero clusters should be the same if $r_i$ contains BCOs. Mathematically, this corresponds to a larger \emph{inner product} of vectors $c_{ij}$ and $1\in \R^{M_i-1}$. Therefore, {step~(4)-e} computes such inner products denoted $\{v_{ij}\}_{j=1}^{D}$ and step~(5) chooses the best.

At the end of step~(5), associated with each subband, we have a real number $b_i$ which characterizes the likeliness of $r_i$ containing BCOs. Finally, for the specified isolation accuracy~$\epsilon$, isolated subband indexes are returned. 

Finally, a potential \texttt{BPM} value is computed as 
\begin{equation}\label{eq:BPM}
\texttt{BPM}=\lceil (1\tran c_{ij}) \rfloor/\texttt{length}(c_{ij})
\end{equation}
for some $i\in \mathcal{I}^\star$, where $\lceil x \rfloor$ represents the rounding of $x$ to the nearest integer and $\texttt{length}(y)$ represents the length of vector $y$.

\section{Computational  Complexity}\label{sec:complexity}

A vast majority of the existing methods use the STFT to obtain a TFR. The asymptotic complexity for the STFT is $\mathcal{O}(N\log{N})$, where $N$ is the samples used in the underlying FFT operations~\footnote{e.g., the STFT window length.}~\cite{cooley1}. 

The discrete S-transform, on the other hand, has an asymptotic complexity of $\mathcal{O}(N^3)$~\cite{wang1}. However, by exploiting structural properties, variants of discrete S-transforms, such as fast discrete orthonormal Stockwell transform can be computed, still in $\mathcal{O}(N\log{N})$~\cite[Theorem~6.1]{wang1}.

\section{Results}\label{sec:results}


This section compares the performance of the proposed method with the algorithms documented in~\cite{ellis1} and~\cite{klapuri4}, which we consider as benchmarks~\emph{A} and~\emph{B}, respectively. Algorithm in~\cite{ellis1} can be considered to be superior among the methods based on pure temporal/spectral analysis methods~\cite{ismir}. On the other hand, the work by~\cite{klapuri4} is the best among methods that rely on temporal/spectral data, together with statistical estimation.

In our simulations, we consider two publicly available datasets - the Ballroom dataset, and the Songs dataset, which comprise of $698$ and $465$ song excerpts, respectively~\cite[\S~III-B]{gouyon1}. The tempo, genre, and style distribution of the datasets are given by~\cite[\S~III]{gouyon1}.

Note that the sampling rate of each song excerpt is $44.1$kHz. A downsampling factor of $D=40$, a subband size $K=1103$, and $Q=10$ subbands are used as inputs to Algorithm~1. In the case of Algorithm~~2, we use $n_p=40$, $H=100$, and $\epsilon=10^{-3}$. 

To exemplify the outputs of the proposed algorithms, we fist consider an arbitrarily chosen classical music excerpt in the Songs dataset.  

\begin{figure}[ht]
	\centering
	\includegraphics[width=0.9\columnwidth]{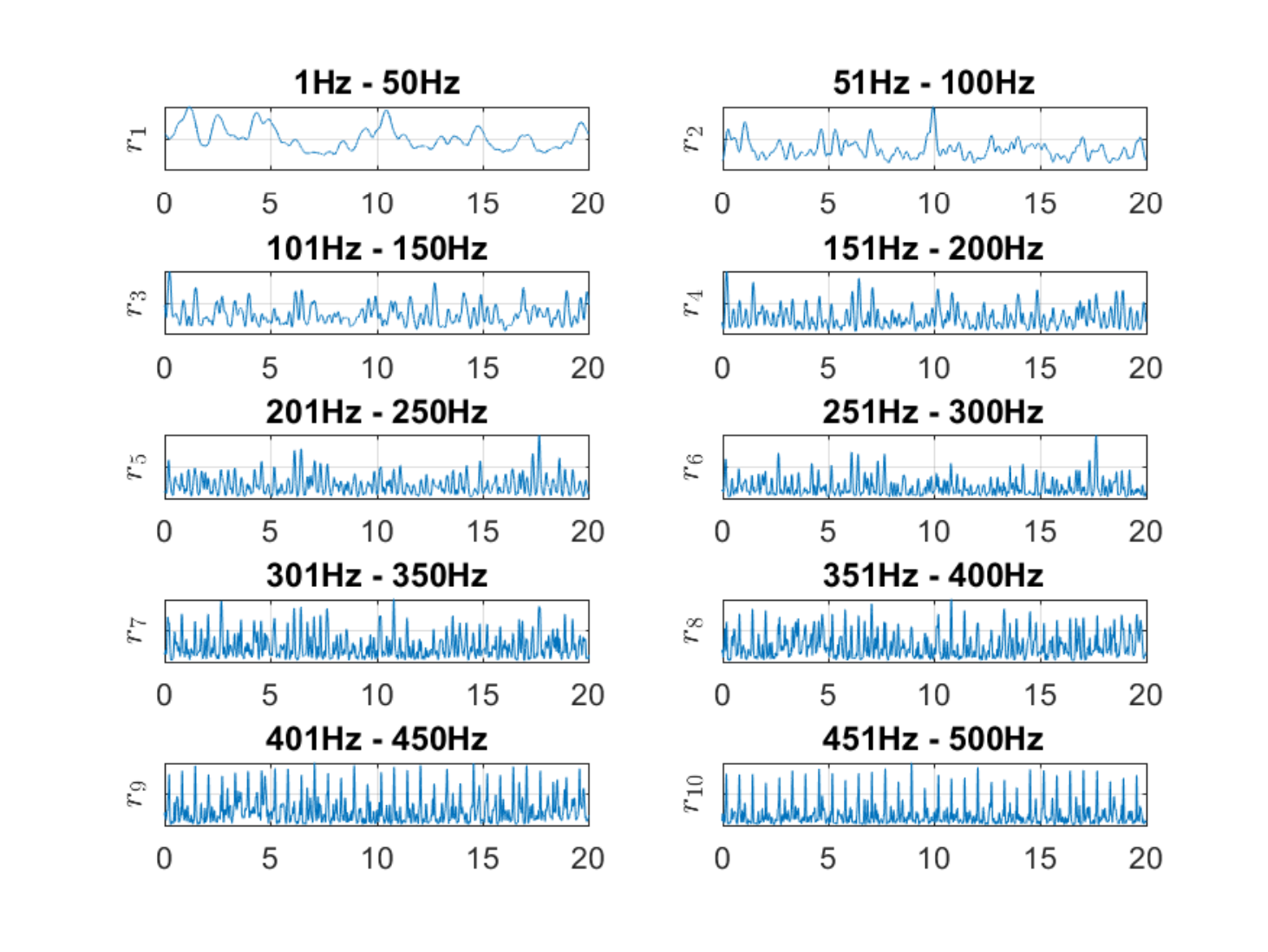}
	\caption{Split frequency bands $r_i$ for $i=1,\ldots,10$}
	\label{fig-bands}	
\end{figure}

Figure~\ref{fig-bands} shows the output $r_i, \ i=1,\ldots, 10$ for the considered music excerpt. Results show that $r_9$ and $r_{10}$ can apparently isolate the BCOs.  

\begin{figure}[ht]
	\centering
	\includegraphics[width=0.9\columnwidth]{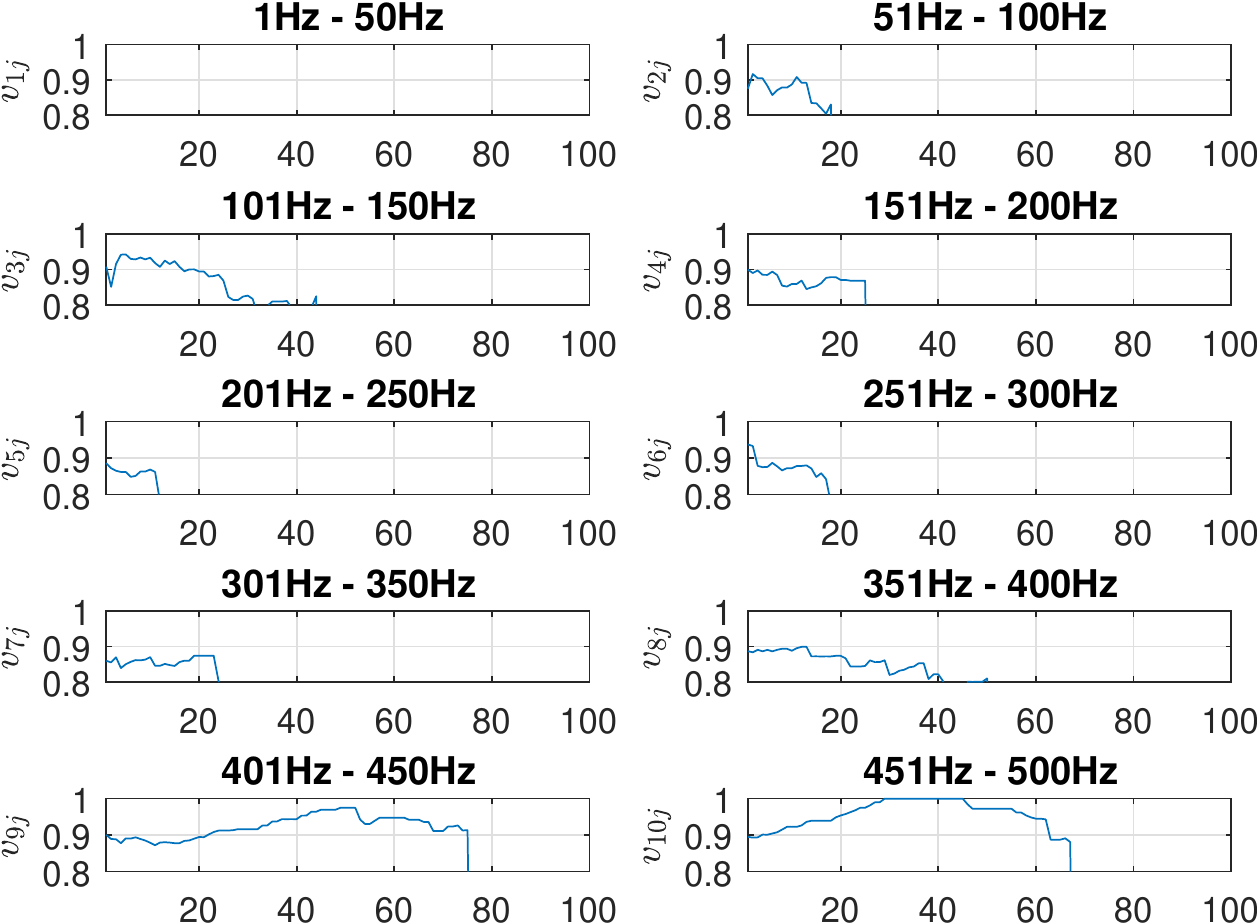}
	\caption{$v_{ij}$ versus $j$ for each subband $i, \ i=1,\ldots,Q$}
	\label{fig-sig}	
\end{figure}

Figure~\ref{fig-sig} shows $v_{ij}$ versus $j$ for each subband $i, \ i=1,\ldots,Q$. Results indicate that $v_{10j}$ yields values almost close to $1$ for some thresholds $l_j$ [\cf step (4)-a]. More specifically, Algorithm~2 returns $\mathcal{I}^\star=\{10\}$, which corresponds to $b_{10}=0.999895$~[\cf step (5)]. The resulting \texttt{BPM} is $88$ [\cf \eqref{eq:BPM}], which is identical to the ground-truth tempo. 

To see the performance of the proposed algorithms on average, we ran the algorithms separately for each data set.

As discussed in~\cite{gouyon1}, we considered the same two metrics to measure the accuracy of the system. In particular, we have
\begin{itemize}
	\item \texttt{Accuracy 1}: The percentage of tempo estimates within 4\% of the ground-truth tempo.
	\item \texttt{Accuracy 2}: The percentage of tempo estimates within 4\% 	of either the ground-truth tempo, or half, double, three times, or one third of the ground-truth tempo.
\end{itemize} 

\begin{figure}[ht]
	\centering
	\includegraphics[width=0.9\columnwidth]{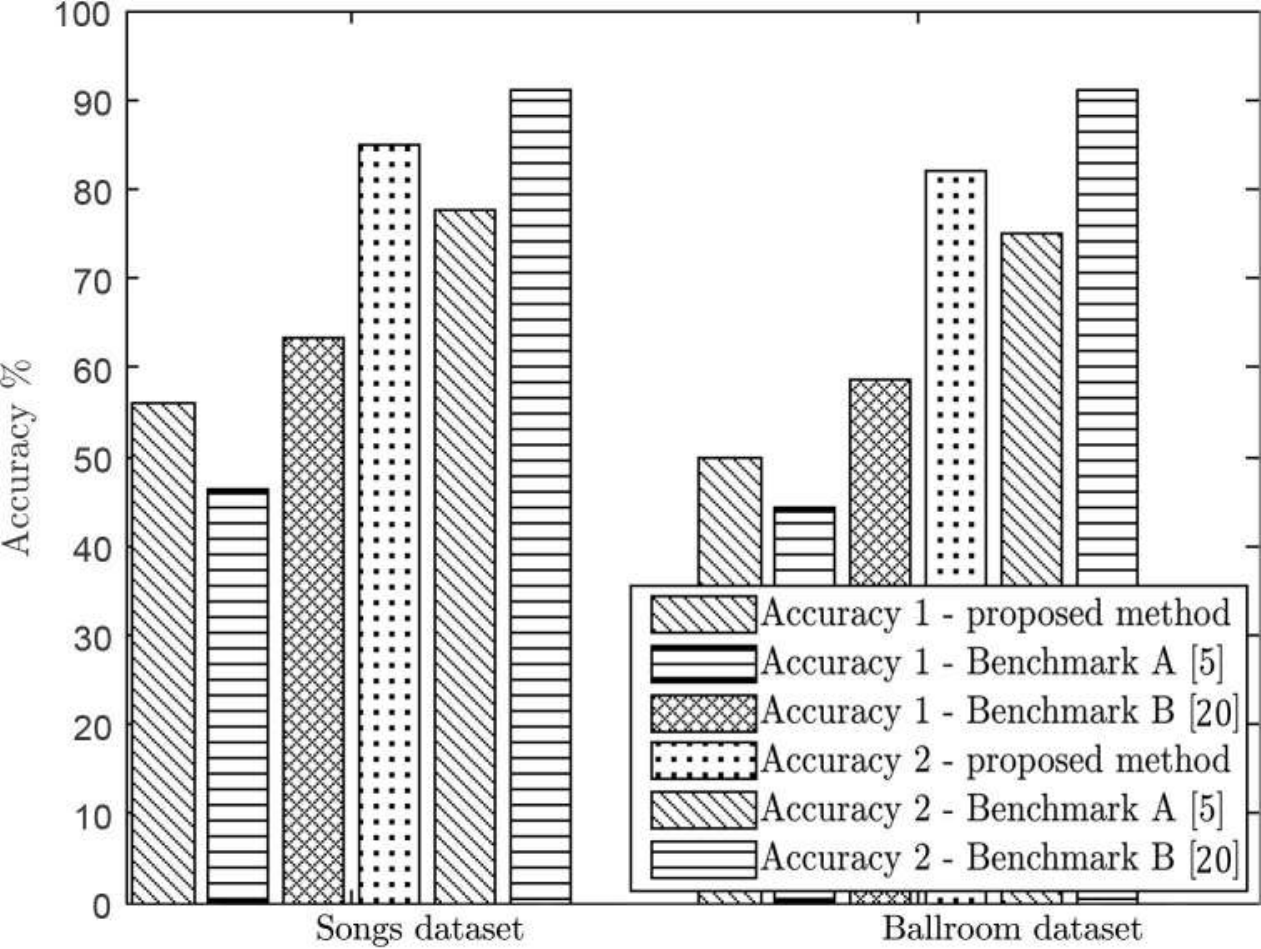}
	\caption{Comparison of Accuracy 1 and Accuracy 2 values for both datasets}
	\label{res-tot}	
\end{figure}

Figure \ref{res-tot} depicts the results of percentage accuracies of the proposed algorithm compared with the two benchmarks. Results show that the proposed method has a better performance than~\cite{ellis1}. Note that, this gain can be accomplished with the same computational complexity, \textit{cf.}~\S~\ref{sec:complexity}. Results further show that the method proposed in~\cite{klapuri4} is superior to the proposed method. This is not surprising, because, unlike the proposed method, the algorithm in~\cite{klapuri4} relies on many computationally intensive operations, e.g., filtering, comb filter operations, discrete power spectral estimations, statistical estimation of period and phase of underlying time series within a hidden Markov model, among others. Therefore, results suggest that the proposed method holds an advantage in that it is less computationally intensive than the benchmark~\cite{klapuri4}, yet with a comparable performance.




\section{Conclusions}\label{sec:Conclusion}

In this paper, a beat causing onset (BCO)  detection method based on the S-transform has been proposed. The method provided an advantage over the approaches that are purely based on classic temporal/spectral analysis. The frequency dependent window dilation used in S-transform has been the key to yield such performances by exploiting better frequency resolution at lower frequencies, where BCOs generally occur. Compared to state-of-the-art algorithms, the proposed method is less resource intensive. For example, our method does not require any \emph{a priori} information of the underlying music, unlike the statistical estimation based approaches. Moreover, the method does not require training datasets like in the methods based on state-of-the-art machine learning techniques. The result is a graceful trade-off between the performance and the required computational burden and the resources.

\bibliography{references}
\bibliographystyle{ieeetr}

\end{document}